# Cooperative lattice theory for $CO_2$ adsorption in diamine-appended metal organic framework at humid direct air capture conditions

Bennett D. Marshall[*], Pavel Kortunov, Aaron Peters, Hilda Vroman

ExxonMobil Technology and Engineering Company, Annandale, NJ 08801, USA.

## Abstract

The effect of humidity on the cooperative adsorption of $CO_2$ from air on amine appended metal organic frameworks is studied both experimentally and theoretically. Breakthrough experiments show that at low relative humidities there is an anomalous induction effect, where the kinetics at short times are slower than kinetics at long times. The induction effect gradually vanishes as relative humidity is increased, corresponding to an increase in $CO_2$ adsorption rate. A new theory is proposed based on the Lattice Kinetic Theory (LKT) which explains these experimental results. LKT is able to accurately represent the measured data over the full range of humidities by postulating that the presence of adsorbed water shifts the equilibrium clusters from cooperatively bound chains to non-cooperatively bound $CO_2$. A consequence of this transition is that $CO_2$ exhibits type V adsorption in dry air, and type I adsorption in humid air.

*bennett.d.marshall@exxonmobil.com

**I: Introduction**

Direct air capture (DAC) is the process of capturing $CO_2$ from the atmosphere using industrial processes.[1] DAC processes are typically either adsorptive[2] or absorptive[3], with most adsorptive processes using supported amine adsorbents. Metal–organic frameworks (MOFs) have commonly been used due to their high porosity, tunability, and periodic array of amine grafting sites.[4, 5] Among the plethora of examples of amine-grafted MOFs, $CO_2$ adsorption in amine-appended $Mg_2$(dobpdc) (dobpdc$^{4-}$ = 4,4′-dioxidobiphenyl-3,3′-dicarboxylate) materials have been well-studied as they exhibit a unique step-like equilibrium behavior that may have some advantages in $CO_2$ capture due to their large swing capacity.[6, 7] The step-like behavior is the result of the cooperative formation of ammonium carbamate chains generated along the crystallographic axis from $CO_2$ insertion into a Mg–N bond and rearrangement of adjacent amines to act as a Bronsted base to aid in charge dissipation.[8]

Siegelman et al.[9] measured breakthrough curves of a synthetic flue gas, $CO_2$ partial pressure of $P_{CO2}$ = 40 mbar, on 2-(aminomethyl)piperidine (2-ampd) appended $Mg_2$(dobpdc). Siegelman considered both humid and dry cases of 40 mbar partial pressure $CO_2$ and demonstrated that humidity had a dramatic effect on $CO_2$ adsorption kinetics and adsorption equilibria. Siegelman et al. concluded that the effect of humidity must either shift the position of the step or change the overall shape of the adsorption isotherm. The presence of humidity drastically improved the performance of the MOF for $CO_2$ capture. Holmes et al.[10] further studied the $CO_2$ loading at a range of humidities in flue gas, and found that there was an optimum humidity which resulted in a maximum $CO_2$ loading.

The work of Siegelman et al. and Holmes et al. were limited to $CO_2$ partial pressures relevant to flue gas capture. The effect of humidity on $CO_2$ adsorption in (2-ampd) appended $Mg_2$(dobpdc) has not been studied at DAC conditions ($P_{CO2}$ = 0.4 mbar). Darunte et al.[11] demonstrated that dry breakthrough curves at DAC conditions exhibited a strong "induction" effect, where the adsorption kinetics where slower at short times than long times. This is opposite the normal behavior, when adsorption kinetics is fastest at short times due to a stronger thermodynamic driving force in unloaded adsorbents. The question then arises, how does this anomalous induction effect depend on relative humidity? Does the presence of water retard the anomalous induction effect, and what are the implications on applicability of this class of materials to $CO_2$ capture processes? To answer these questions, experiments must be performed at low $CO_2$ partial pressures (0.4 mbar) and a wide range of relative humidities. In addition, a theoretical construct is required to interpret the experimental data and make predictions based on the analysis of this data.

Adsorption of $CO_2$ in diamine-appended $Mg_2$(dobpdc) and $Mn_2$(dobpdc) occurs in 6 lanes which act independently.[12] For this reason, adsorption occurs in an approximately 1-dimension(1-D) space. This dimensionality, combined with the fact that adsorption occurs at specific lattice sites, invites the application of a statistical mechanics based lattice theory. Kundu et al.[12] developed a cooperative lattice theory for adsorption in these diamine-appended MOFs by considering different states of adsorbed $CO_2$: non-cooperatively adsorbed, cooperatively adsorbed chain ends, and cooperative adsorbed chain interiors. Non-cooperatively (singly) adsorbed $CO_2$ interacts with a single amine / metal lattice site, while cooperatively adsorbed $CO_2$ requires the interaction with two amine / metal lattice sites. Non-cooperatively bound $CO_2$ species may be on adjacent lattice sites; however the two adsorbed sites not directly interact. It is the cooperative chains which give rise the step isotherms. The lattice theory of Kundu et al. can be accurately applied as a methodology to regress adsorption data; however, unlike the purely empirical isotherm methodologies, the regression of the lattice theory extracts physical information on the adsorbed phase, such as the fraction of $CO_2$ adsorbed non-cooperatively and cooperatively.

The additional structural information provided by the lattice theory allowed Marshall[13] to construct a kinetic theory, lattice kinetic theory (LKT), based on the morphology of the adsorbed / reacted $CO_2$. Non-cooperatively bound and cooperatively bound $CO_2$ were allowed separate but coupled rate laws, which allowed for the prediction of the anomalous "induction effect" observed in the low partial pressure breakthrough curves of Darunte et al.[11] and the TGA assays of Martell et al.[14] This induction effect was explained by LKT due to an asymmetry in adsorption rate laws between cooperative and non-cooperatively bound $CO_2$.

Given the success of applying LKT to breakthrough curves of dry 0.4 mbar partial pressure $CO_2$ streams, we now wish to apply LKT to model humid breakthrough data. Can the humidity induced enhancement of $CO_2$ equilibrium adsorption and kinetics be explained on the basis of a shift in the morphology of equilibrium adsorbed clusters?

The paper is organized as follows. Section I discuss the synthesis and characterization of the test material 2-ampd appended $M_2(dobpdc)$, where M is 99.4% Mg and 0.6% Mn. Section III introduces the experimental methodology to measure humid breakthrough curves at DAC conditions. In section IV, LKT is extended to double step isotherms and is applied to dry equilibrium and kinetic adsorption data for 2-ampd appended $M_2(dobpdc)$. In section V, the theory is extended to humid air, and applied to explain the observed experimental breakthrough curves.

**II: Material synthesis and characterization**

$Mn_{0.12}Mg_{1.88}$(dobpdc) was prepared as published previously.[15] 2-(aminomethyl)piperidine (2-ampd) for diamine grafting was purchased from ChemShuttle. To append the amine to the MOF, the MOF was first solvent-exchanged to ethanol by soaking roughly 5 g of wet cake in 40 mL ethanol. The material was then centrifuged (5000 rpm, 5 min) and the ethanol was decanted and replaced with 40 mL of fresh ethanol. This solvent exchange procedure was repeated three additional times. The material was then placed in a vacuum oven at 70 °C for four hours to dry the material sufficiently. Roughly 1 g of dried MOF was then added to a 20-mL vial to which 20.4 g of toluene and 0.36 g of 2-ampd was added. The MOF/amine slurry was stirred overnight at room temperature. The amine-appended MOF was then centrifuged (5000 rpm, 5 min) and the excess toluene/amine solution was decanted. The MOF was then washed five additional times with fresh toluene (20 g) and finally dried at 70 °C in a vacuum oven. A series of characterization tests were conducted to assess the quality of the material, and are discussed in the Appendix.

Note, the formulation deviates very slightly from Siegelman et al.[9], who employed a pure magnesium metal. However, the low level of Mn used here does not significantly effect the physical properties or performance of the material. This is demonstrated in Fig 3., which shows isotherms measured by Siegelman et al.[9] using the pure magnesium material, and isotherms measured on the current material $Mn_{0.12}Mg_{1.88}$(dobpdc) are consistent.

**III: Breakthrough measurements**

Column gas breakthrough measurements were performed to assess the effect of humidity on the equilibrium and kinetics of CO2 adsorption of the prepared MOF. Breakthrough measurements are defined by having a column packed with adsorbent material, and flowing gas containing a known composition of adsorbing species ($CO_2$ + $H_2O$) in an inert carrier gas through the column. The composition of the gas exiting the column is measured as a function of time, allowing for quantification of adsorption kinetics as well as equilibrium capacity. Breakthrough test probe the adsorption kinetics over the full range of compositions up to and including the composition of the feed gas. A diagram of the breakthrough unit can be found in Fig. 1.

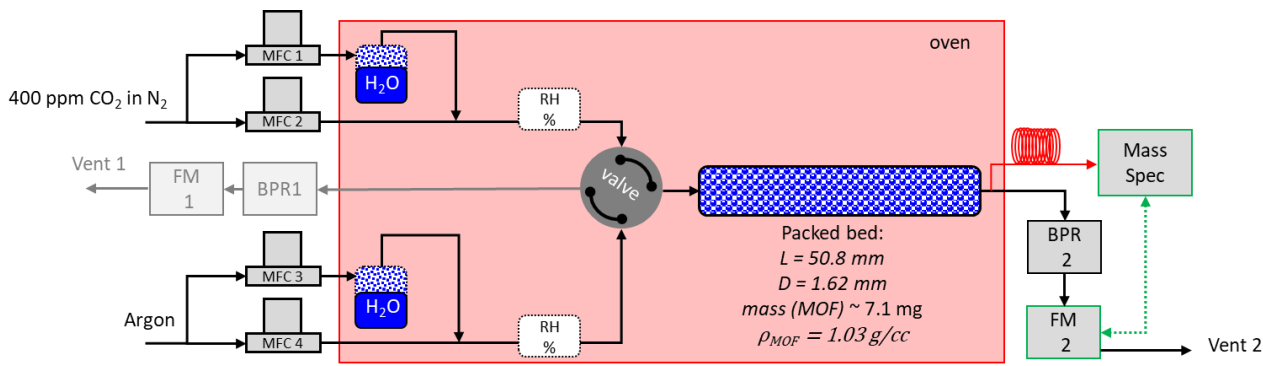

**Figure 1:** Diagram of column breakthrough process

A total of $m_{MOF}$ = 7.1 mg of MOF crystals agglomerates sieved to approximately 56 micrometers were packed into a column of length $L$ = 50.8 mm and diameter $D$ = 1.63 mm. The crystal density of MOF of $\rho_{MOF}$ = 1.03 g / cc was used to calculate the bed porosity $\rho = 1 - m_{MOF}/\rho_{MOF}V$, where $V$ is the bed volume. Humid and dry feed streams containing CO2 were blended to achieve a desired humidity as measured by calibrated humidity meter before the column. $CO_2$ concentration in certified gas mixture was maintained at 400 ppm on dry gas basis and verified by calibrated gas analyzer. Dry argon was used as a purge gas to establish a stable gas flow through the column before CO2 breakthrough experiment and to activate the column after CO2 adsorption experiment. Column off gas analysis was performed with a HIDEN mass spectrometer calibrated before and after every experiment. The inlet gas flowrate was maintained at 10.7 SCCM while outlet flow rate was directly measured by calibrated OMEGA flowmeter. Calculations of $CO_2$ and $H_2O$ adsorption were based on mass balance via integration of inlet and outlet gas compositions and flow rates over time. Total adsorption of CO2 and water were additionally verified via weighing the column before and after the breakthrough experiment. Column with MOF was activated by purging dry argon gas at 120C after each experiment to desorb previously adsorbed CO2 and water molecules and regenerate MOF crystals. After regeneration, the column was cooled down to target temperature of 22C in dry argon purge to prepare the column for next experiment at different humidity of the CO2 containing feed gas.

**IV: Theory and Analysis for dry direct air capture conditions**

In this section we extend the lattice kinetic theory (LKT)[13] to the case of $CO_2$ adsorption on 2-ampd appended $M_2$(dobpdc) at dry direct air capture conditions. In addition to theory development, we include the application of the theory to measured data of $CO_2$ adsorption from dry air in this section.

LKT is built around the cooperative 1D lattice theory of Kundu et al.[12], which accounts for several different states of adsorbed $CO_2$. Figure 2 illustrates the three adsorbed states. $CO_2$ may exist in cooperative chains or non-cooperative singles. Within the cooperative chains, $CO_2$ can exist either as a chain end or chain interior. Likewise, non-cooperatively bound singles may exist on neighboring lattice sites; however, the non-cooperative singles only interact with the lattice site which they occupy, not filled neighboring lattice sites.

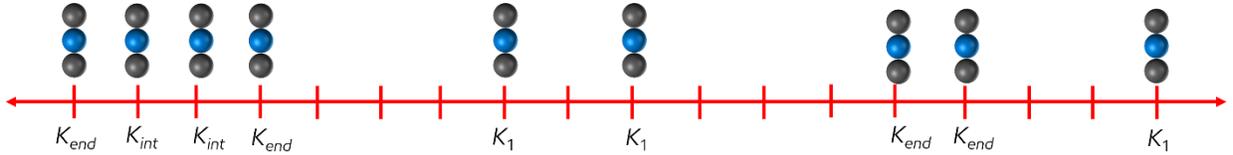

**Figure 2:** Diagram of adsorbed $CO_2$ on 1D lattice

Each type of adsorbed $CO_2$ is represented by an equilibrium constant. $K_1$, $K_{end}$ and $K_{int}$ are the equilibrium constant of non-cooperatively bound singles, cooperative chain ends, and cooperative chain interiors respectively. For an ideal gas the equilibrium constants are calculated theoretically through the relation

$$K_\alpha = \frac{V_\alpha P_{CO_2}}{k_B T} \exp\left(\frac{E_\alpha}{k_B T}\right) \qquad \alpha = 1, int, end \qquad (1)$$

where $P_{CO_2}$ is the partial pressure of $CO_2$, $V_\alpha$ is the entropic volume for adsorbed state $\alpha$, $E_\alpha$ is the lattice interaction energy for adsorbed state $\alpha$, and $k_B$ is Boltzmann's constant.

The lattice theory has a strong theoretical foundation based in statistical thermodynamics. The grand free energy $f$ contains all equilibrium information of the adsorbed phase.[12]

$$f = -k_B T \ln \lambda_+ \tag{2}$$

$\lambda_+$ is the maximum Eigen value of the transfer matrix.

$$2\lambda_+ = 1 + K_1 + K_{int} + \sqrt{(1 - K_1 - K_{int})^2 + 4K_{end}^2} \tag{3}$$

The dimensionless density $\rho$ of adsorbed $CO_2$ is obtained from the partial pressure derivative of Eq. (2)

$$\rho = \frac{P_{CO_2}}{\lambda_+} \frac{\partial \lambda_+}{\partial P_{CO_2}} \tag{4}$$

The adsorption in terms of moles of adsorbed $CO_2$ per mass of adsorbent is obtained by multiplication of Eq. (4) by the maximum adsorbent loading $q_o$.

$$q = q_o \rho = q_o \frac{P_{CO_2}}{\lambda_+} \frac{\partial \lambda_+}{\partial P_{CO_2}} \tag{5}$$

A major benefit of using the statistical mechanics based theory, as compared to a purely empirical correlation, is that structural details of the adsorbed phase can be extracted from appropriate derivatives of the grand free energy. The equilibrium adsorption of non-cooperative singles, chain ends and chain interiors are of central importance to the kinetic theory. These values are calculated via derivatives as described in the SI of Kundu et al. with $\omega = 1 / \lambda_+$

$$\frac{q_1}{q_o} = \frac{K_1(1-\omega K_{int})^2}{(1+K_1)(1-\omega K_{int})^2 + K_{end}^2 \omega(2-\omega K_{int})} \quad (6)$$

$$\frac{q_{end}}{q_o} = \frac{2K_{end}^2 \omega(1-\omega K_{int})}{(1+K_1)(1-\omega K_{int})^2 + K_{end}^2 \omega(2-\omega K_{int})} \quad (7)$$

$$\frac{q_{int}}{q_o} = \frac{K_{end}^2 K_{int} \omega^2}{(1+K_1)(1-\omega K_{int})^2 + K_{end}^2 \omega(2-\omega K_{int})} \quad (8)$$

The relative magnitude of the equilibrium constants $K_\alpha$ controls the shape of the isotherm. For $K_1 \gg K_{int}, K_{end}$ the theory will produce a Langmuir like type I isotherm, due to the dominance of singly adsorbed $CO_2$ clusters. For both $K_{int}, K_{end} \gg K_1$ chains will be favored, resulting in a step like isotherm. For step isotherms, if $K_{end} \sim K_{int}$ neither chain ends nor interiors are favored. This results in broad steps in the isotherm. On the other-hand, for $K_{end} < K_{int}$, chain interiors are favored over chain ends. The system will then act to minimize chain ends, resulting in long adsorbed chain clusters and a sharp step in the isotherm.

The lattice theory is described by a total of 7 parameters, the $V_\alpha$ and $E_\alpha$ of each adsorbed state, as well as the maximum loading $q_o$. These parameters must be adjusted to measured adsorption data. Marshall[13] demonstrated the theory can be accurately fit to represent sorption data on mmen-$Mg_2$(dobpdc) using only 4 adjustable parameters, by assuming $V_1 = V_{end} = V_{int}$ and $E_1 = E_{end}$.

$CO_2$ exhibits a double step on 2-ampd-$M_2$(dobpdc).[9] To represent the double step using the lattice theory we must take the total adsorption as a sum of two lattice theories

$$q = q^{(1)} + q^{(2)} \quad (9)$$

where $q^{(1)}$ is the adsorption on the first lattice and $q^{(2)}$ is the adsorption on the second. Likewise, the lattice interaction volumes, energies as well as maximum loading are expressed as $V_\alpha^{(k)}$, $E_\alpha^{(k)}$ and $q_o^{(k)}$, where the superscript $k$ refers to lattice 1 or 2. We place the following restrictions on the parameters

$$q_o^{(1)} = q_o^{(2)} \tag{10}$$

$$E_1^{(1)} = E_1^{(2)} \tag{11}$$

$$V_1^{(1)} = V_1^{(2)} \tag{12}$$

Equation (10) enforces that the maximum capacity of each lattice are identical while Eqns. (11) – (12) enforce the equivalence of non-cooperatively bound singles in both lattices. The two lattices differ in the chain parameters. We further restrict the entropic volumes of chain formation such that,

$$V_{ch}^{(k)} = V_{end}^{(k)} = V_{int}^{(k)} \tag{13}$$

In total, that leaves 9 parameters which are adjusted to experimental adsorption data. The model fits are compared to adsorption data in Fig. 3, and the theory parameters are given in Table 1. As can be seen, the theory accurately fits the data. In the bottom panel of Fig. 3 we include theory predictions of the fraction of adsorbed species which exist as non-cooperative singles, cooperative chain ends and cooperative chain interiors at low $CO_2$ partial pressures and $T = 22$ °C. In the range of 0 – 0.2 mbar $P_{CO_2}$ the adsorbed phase transitions from being dominated by non-cooperative singles to being dominated by cooperative chain interiors. In the course of a direct air capture process this structural transition would be occurring through the adsorption system.

**Table 1:** Lattice parameters for adsorption of $CO_2$ on 2-ampd-$M_2$(dobpdc)

| Step | $q_o$(mol/kg) | $V_1$(Å$^3$) | $V_{ch}$(Å$^3$) | $E_1$(kj/mol) | $E_{end}$(kj/mol) | $E_{int}$(kj/mol) |
|------|---------------|--------------|-----------------|---------------|-------------------|-------------------|
| 1    | 1.7           | 1.32x10$^{-6}$ | 1.36x10$^{-5}$ | 65            | 68                | 75                |
| 2    | 1.7           | 1.32x10$^{-6}$ | 6.44x10$^{-5}$ | 65            | 61.75             | 67                |

For both steps 1 and 2, $E_{int} > E_{end}$ which results in step isotherms. The step is a result of chain interiors being favored over chain ends and singles. For this reason, the system creates long cooperative chains. At partial pressures below the step, the entropic penalty of chain formation is too large, while at partial pressures above the step, the enthalpic benefit of chain formation outweighs the entropic penalty.

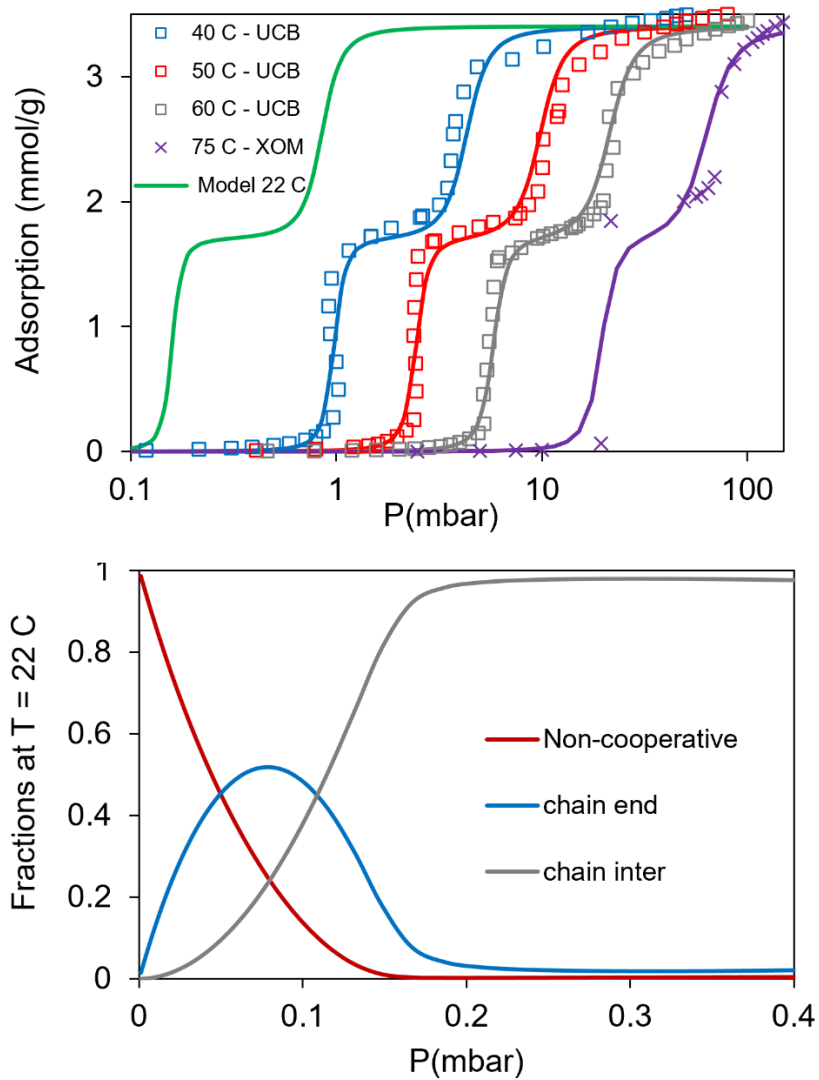

**Figure 3:** Top: Experimental data (symbols) and theory fits (curves) for the adsorption of $CO_2$ on 2-ampd-$M_2$(dobpdc). Squares represent data measured by Siegelman et al.[9] and crosses represent data measured in this study. Bottom: Theory predictions of the fraction of adsorbed species which exist as non-cooperative singles, cooperative chain ends and cooperative chain interiors at low $CO_2$ partial pressures and $T = 22$ °C.

We now focus on the kinetics of $CO_2$ adsorption by employing the recently developed lattice kinetic theory (LKT)[13]. LKT applies different adsorption rates for $CO_2$ adsorbing into non-cooperative singles and cooperative chains. For each step we have ($k = 1,2$)

$$\frac{dq^{(k)}}{dt} = \frac{dq_1^{(k)}}{dt} + \frac{dq_{ch}^{(k)}}{dt} \tag{14}$$

where $q_{ch}^{(k)}$ is the $CO_2$ loading in chains

$$q_{ch}^{(k)} = q_{end}^{(k)} + q_{int}^{(k)} \tag{15}$$

The rate of adsorption of non-cooperatively bound singles is given by a standard linear driving force relation[16]

$$\frac{dq_1^{(k)}}{dt} = k_1 \left( q_{1,eq}^{(k)} - q_1^{(k)} \right) \tag{16}$$

where $q_{1,eq}^{(k)}$ represents the equilibrium single adsorption and the $k_1$ is an adjustable rate constant. We assume $k_1$ is the same for each step. LKT assumes a kinetic mechanism in which chain formation must be first seeded by the adsorption of non-cooperative singles. Chains then grow by $CO_2$ adsorbing at adjacent sites and cooperating with other $CO_2$ molecules at neighboring sites. The rate of chain formation is given by a linear driving force with respect to adsorption of cooperative chains, but is mediated by the ratio of at singles adsorption to the equilibrium singles adsorption.

$$\frac{dq_{ch}^{(k)}}{dt} = k_{ch} \left( q_{ch,eq}^{(k)} - q_{ch}^{(k)} \right) \frac{q_1^{(k)}}{q_{1,eq}^{(k)}} \tag{17}$$

We assume $k_{ch}$ is the same for each step. It is the ratio $q_1^{(k)} / q_{1,eq}^{(k)}$ which enforces that singles must first adsorb before chains can form. If single formation is infinitely slow then $q_1^{(k)} = 0$, and the rate of chain formation will be $dq_{ch}^{(k)}/dt = 0$. On the other hand, if single formation is very fast, then $q_1^{(k)} / q_{1,eq}^{(k)} = 1$, and Eq. (17) takes on the form of a standard linear driving force theory. It is this coupling of the rate of chain adsorption to singles adsorption which gives rise to the observed[9,11] induction behavior.

For adsorption of $CO_2$ in the absence of humidity, the kinetic theory consists of the two constants $k_1$ and $k_{ch}$, which are adjusted to the measured breakthrough data. The breakthrough unit is modelled using mass balance, momentum balance, and LKT for adsorption kinetics. The detailed transport equations are listed in our previous[13] publication. As described in Section III, at the beginning of each experiment the bed is completely devoid of both $CO_2$ and water. The fit rate constants are given in Table 2. $k_1$ is more than an order of magnitude larger than $k_{ch}$, meaning non-cooperative singles formation is much faster than cooperative chain formation. This result is consistent with the proposed kinetic mechanism. Singles dominate at low partial pressures when $CO_2$ loading in the MOF is small. When a single $CO_2$ molecule enters the MOF in this low partial pressure region, there is an abundance of available adsorption sites. This site availability results in faster kinetics. On the other hand, chains dominate at higher partial pressures when the MOF is substantially loaded with $CO_2$. In the assumed mechanism chains grow by adsorption on the chain ends. In a loaded MOF, chain ends will be rare, therefore decreasing the rate at which adsorbing $CO_2$ can adsorb into chains.

**Table 2:** LKT rate constants

| $k_1$ | $k_{ch}$ |
|---|---|
| $3.6 \times 10^{-3}$ s$^{-1}$ | $1.8 \times 10^{-4}$ s$^{-1}$ |

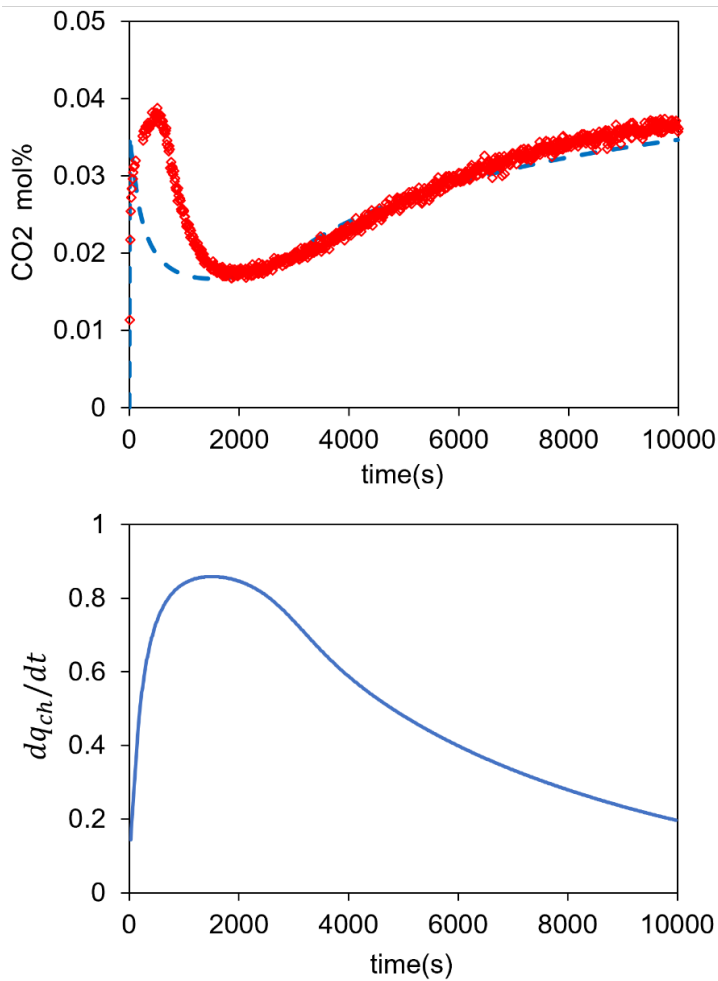

**Figure 4:** Top: Comparison of theory (curve) to data(symbols) for the breakthrough curve of 400 ppm $CO_2$ and a flowrate of 10 SCCM. Bottom: Theory predictions of the time dependence of the rate of chain formation.

The top panel of Fig. 4 compares theory and experiment for the breakthrough curve. The theory is in good agreement with the data for t > 2000 s, but loses accuracy at shorter times. This inaccuracy at small times may be due to inaccuracies in the equilibrium lattice theory. The breakthrough curves are measured at 22 °C, which is a significant extrapolation from the measured equilibrium data to which the lattice theory was fit. Both theory and experiment exhibit the induction effect where the outlet composition goes through a minimum. The bottom panel of Fig.

4 shows the time dependence of the rate of chain formation. This rate is initially zero and then increases, going through a maximum around 1500 s and then decreases. It is this initial delay in adsorption in chains which gives rise to the observed induction effect.

**V: Theory and analysis for humid direct air capture conditions**

Humidity has a profound effect on $CO_2$ adsorption capacity and kinetics for 2-ampd-$M_2$(dobpdc).[9, 10] In this section we modify LKT to account for the co-adsorption of water. We begin with the correlation of the pure water adsorption data of Siegelman et al.[9] at 30 °C. The correlation is a function of relative humidity ($0 \leq RH \leq 1$)

$$q_w = q_{w,1}^{(0)} \frac{(b_1 RH)^{n_1}}{1+(b_1 RH)^{n_1}} + q_{w,2}^{(0)} \frac{(b_2 RH)^{n_2}}{1+(b_2 RH)^{n_2}} \tag{18}$$

with parameters given in Table 3

**Table 3:** Water isotherm parameters. Loadings are in units of mmol / g

| $q_{w,1}^{(0)}$ | $b_1$ | $n_1$ | $q_{w,2}^{(0)}$ | $b_2$ | $n_2$ |
|---|---|---|---|---|---|
| 4.5 | 2.5 | 2 | 11 | 2.1 | 20 |

Figure 5 compares correlation fits to the data of Siegelman et al.[9] as well as water adsorption points measured in this study from breakthrough curves, for $CO_2$ free and in the presence of 400 ppm of $CO_2$.

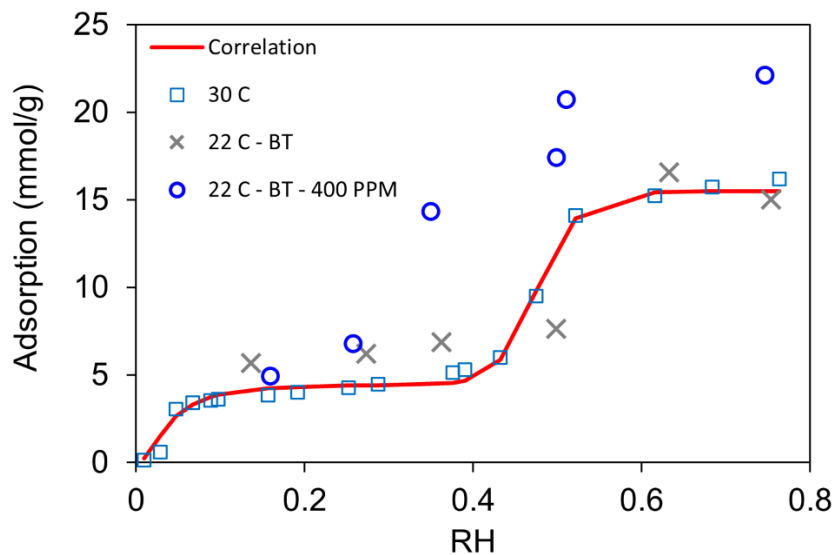

**Figure 5:** Correlation fit (curve) to the pure water adsorption data of Siegelman et al.[9] (squares) versus relative humidity. Crosses and circles give current water adsorption measurements via breakthrough curves. Crosses are measurements in the absence of $CO_2$, while circles are in the presence of 400 ppm $CO_2$.

The developed correlation accurately represents the adsorption data of Siegelman et al. The crosses represent the adsorption of water measured via breakthrough curve in absence of $CO_2$. The breakthrough data shows higher water adsorption at low relative humidities than Siegelman et al., and a pore filling step which occurs at higher relative humidities. It is common that breakthrough measurements of water capacity are not entirely consistent with gravimetric measurements.[17] The blue circles in Fig. 5 represent water adsorption from humid nitrogen in the presence of 400 ppm $CO_2$. As can be seen, the presence of adsorbed $CO_2$ results in a substantial increase in water co-adsorption. This is likely the result of fact that adsorbed $CO_2$ is an ionic form, therefore making the MOF a more polar adsorbent for water.

The presence of humidity increases both $CO_2$ adsorption capacity and kinetics. Here we propose a simple method to correct both adsorption capacity and kinetics by having the presence of adsorbed water affect the adsorbed phase structure. The justification is as follows. The observed step isotherm in dry cases is a result of the formation of carbamate chains along the crystallographic axis.[6] These chains represent a low energy and highly ordered low entropy state. This order results from the charge balancing of neighboring carbamate amine anions and protonated amine cations.

The force between a cation (charge $q$) and an anion (charge $-q$) separated by a distance $r$ is given by Coulombs law.

$$F = -\frac{q^2}{4\pi\varepsilon\varepsilon_o r^2} \tag{19}$$

$\varepsilon$ is the dielectric constant, and $\varepsilon_o$ is the vacuum permittivity. In the absence of co-adsorbed water, the space between the anion and cation will be vacuum, with a dielectric constant of $\varepsilon = 1$. However, in a water loaded MOF, the dielectric constant will be like that of water, $\varepsilon \sim 80$. This means, that the attractive force between a carbamate anion, and a protonated amine cation, will be nearly 80 times smaller in the presence of co-adsorbed water, as compared to dry case. As a result of this charge screening, the driving force to create the highly ordered low entropy carbamate chains will be substantially decreased in the presence of co-adsorbed water. In turn, this will decrease the cooperative nature of the adsorption which gives rise to step isotherms.

In the context of LKT, this postulated mechanism can be represented by having co-adsorbed water transition the adsorbed $CO_2$ clusters from cooperative chain dominated in the absence of water, to non-cooperative singles dominated in the presence of water. Further supporting this proposition is the fact that the rate constant $k_1$ is a factor of 20 larger than $k_{ch}$, Table 2. Therefore, it may be possible to represent both the increase in equilibrium adsorption, as well as the increase in kinetics, by enforcing that co-adsorbed water results in a transition from cooperative chain dominated to non-cooperative single dominated adsorption equilibria.

To achieve this, we will rescale the entropic volume $V_1$ as a function of co-adsorbed water $q_w$.

$$Vs = \frac{V_1(q_w)}{V_1(0)} \quad (20)$$

In Eq. (20), $V_1(q_w)$ is the non-cooperative volume at co-adsorbed water loading $q_w$ (Eq. 18) and $V_1(0)$ is the dry volume given in Table 1. Figure 6 gives theory predictions of the fraction of cooperative and non-cooperative clusters, as well as adsorption isotherm for several values of $Vs$. The curves for $Vs = 1$ correspond to the dry calculations in Fig 3. As $Vs$ is increased, the two steps in the adsorption isotherm are deformed, becoming less step like. For $Vs = 500$, the steps have disappeared altogether, resulting in Type I adsorption behavior. From the top panel in Fig. 6, it is clear the transition from cooperative chains to non-cooperative singles is driving this change.

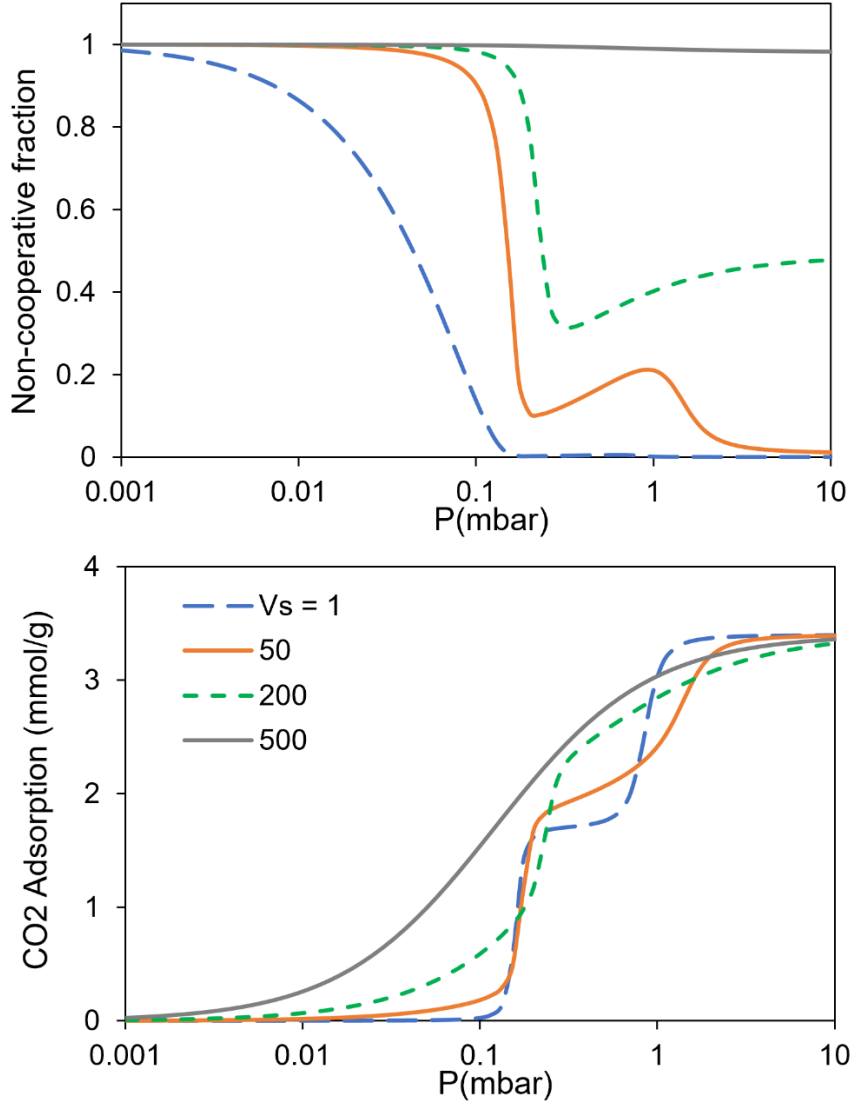

**Figure 6:** Top: Theory predictions of the fraction of adsorbed clusters which are non-cooperative singles at T = 22 °C for several scaling factors *Vs* defined in Eq. (20). Bottom: Theory predicted adsorption.

We propose the following simple correlation for *Vs*

$$\frac{V_1(q_w)}{V_1(0)} = Vs = 1 + c_1 \frac{q_{w,1}}{q_{w,1}^{(o)}} + c_2 \frac{q_{w,2}}{q_{w,2}^{(o)}} \tag{21}$$

where $q_{w,1}$ and $q_{w,2}$ are the equilibrium adsorption of the first and second water steps in Fig. 5 and $q_{w,1}^{(o)}$ and $q_{w,2}^{(o)}$ are the maximum equilibrium loadings of these steps given in Table 3. The $c_1$ and $c_2$ are empirical constants which are adjusted to reproduce the breakthrough measurements and shown in Table 4. As can be seen, $c_2 = 1200$ is much larger than $c_1 = 50$, implying that the pore filling transition at high relative humidities has a much stronger effect on $CO_2$ adsorption than the site-based water adsorption observed at lower relative humidities.

**Table 4:** Rescaling constants in Eq. (21)

| $c_1$ | $c_2$ |
|---|---|
| 50 | 1200 |

Equation (21) completes the theory. For the remainder of this section, we compare theory predictions to experimental breakthrough curve measurements of 400 ppm $CO_2$ in humid nitrogen. Figure 7 compares theory and experiment for the time dependent adsorption during the breakthrough measurement. Increasing humidity results in an increase in both equilibrium capacity and kinetics. The theory overpredicts adsorption at RH = 25%, and underpredicts at RH = 36%; however, the overall theory agreement with experiment is satisfactory.

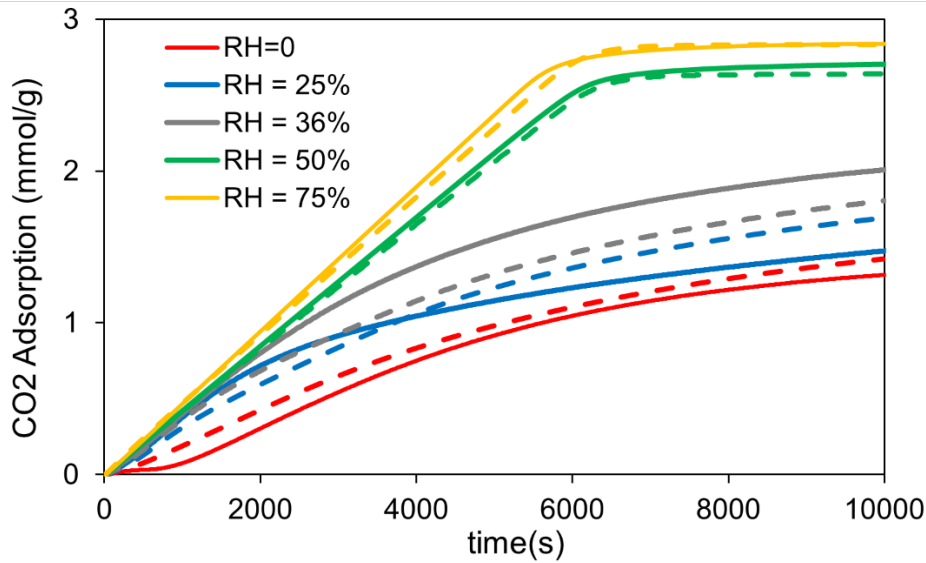

**Figure 7:** Comparison of theory (dashed curves) to experiment (solid curves) for $CO_2$ adsorption during breakthrough cures at several relative humidities

Figure 8 compares theory and experiment for the breakthrough curves. Comparing the moderate humidity cases (16%, 25%, 36%) we see that much less $CO_2$ slips through the bed at short times (t < 2000 s) as compared to the dry cases. The outlet composition goes through a minimum in these moderate humidity cases for t < 2000 s and then begins to increase. From Fig. 4 we see that each of these moderate humidity cases occurs at humidities below the water pore filling transition in the pure water case. The high humidity cases (50%, 75%) occur above the pore filling transition of water. The high humidity breakthrough curves are qualitatively different those of moderate humidity. At high humidity the breakthrough curves are sharp, characterized[11] as a "shock" breakthrough. The experimental breakthrough curves measured at high humidity exhibit a small slip which is not reproduced in the theory. This may be due to diffusional limitations resulting from the requirement for $CO_2$ to diffuse through liquid like water in the center of the MOF crystals.

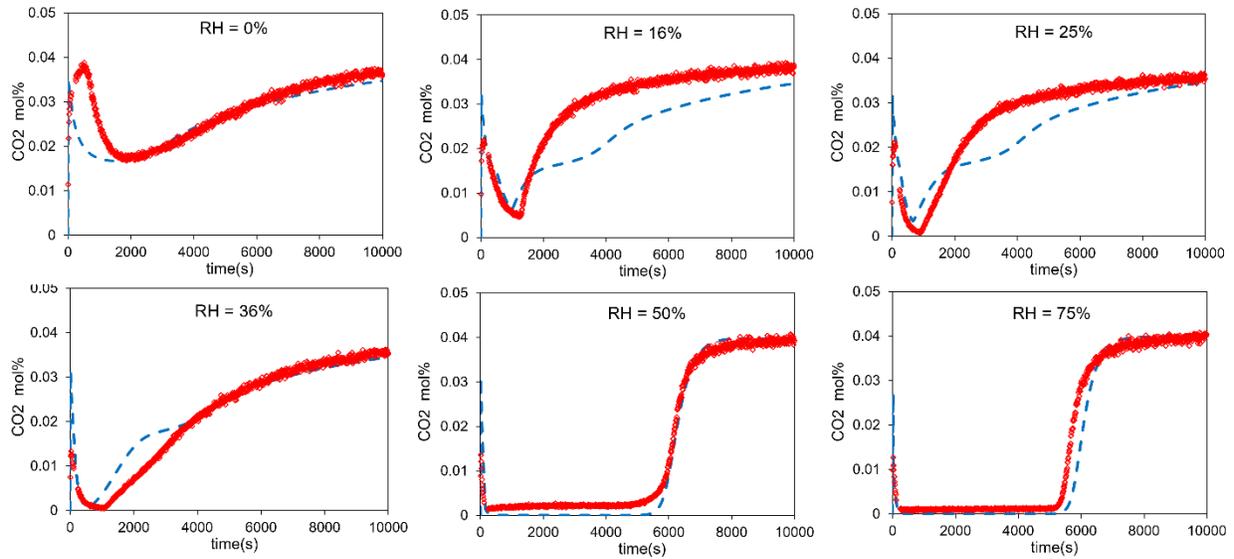

**Figure 8:** Comparison of theory (dashed curve) and experiment (red circles) for the breakthrough curves at several relative humidities

From Fig. 8, it is shown that at dry conditions the breakthrough curve exhibits an induction effect, where kinetics at short times is slower than at long times. This is opposite the normal adsorption behavior, where short time kinetics are faster due to a higher driving force for adsorption. As relative humidity is increased, this induction effect disappears altogether. LKT gives a clear physical explanation for this process. Figure 9 shows theory predictions of the fraction of adsorbed $CO_2$ which exist as non-cooperative singles during the breakthrough test. For each humidity, non-cooperative singles dominate at very short times. For humidities up to 36%, cooperative chains dominate at long times. For the high humidity cases (50%, 75%) the adsorbed phase is nearly 100% non-cooperative singles at long times. Figures 7 and 8 show that breakthrough curves which exhibit sharp shock breakthroughs are dominated by non-cooperative singles, while cooperative chain dominated systems (RH ≤ 36%) exhibit minima in the breakthrough curves.

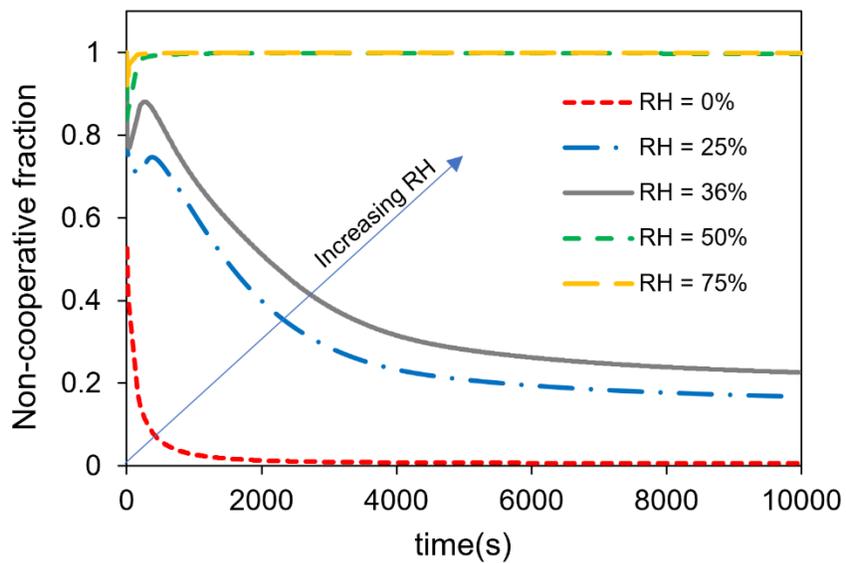

**Figure 9:** Theory predictions of fraction of adsorbed species which exist as non-cooperative singles during breakthrough

**VI: Conclusions**

The combined experimental and theoretical analyses presented in this work allowed elucidation of the effect of humidity on both the equilibrium and kinetic adsorption of dilute $CO_2$ on 2-ampd appended $M_2$(dobpdc). It was shown that both the enhancement in equilibrium adsorption and adsorption rate with increasing humidity could be described by the transitioning from cooperative to non-cooperative adsorbed clusters in the presence of humidity. On this basis, the lattice kinetic theory was extended to humid air, and was shown to accurately represent the breakthrough data. A significant prediction of the theory is that the step isotherm observed in dry conditions disappears in humid conditions. This has significant ramifications on the design and modelling of direct air capture processes.

**Appendix:**

A series of characterization tests were conducted to assess the quality of the material. First, the material was digested in an NMR solvent (two drops of DCl in 0.5 mL d6-DMSO). Figure A1 shows the $^1$H-NMR spectrum of the resulting digested material. Integrating peaks associated with the linker (7 ppm, 7.5–8 ppm) and the amine (1.25–2 ppm, 2.75–3.5 ppm) indicates that the material is 100% loaded (i.e. one amine molecule per metal site). A $CO_2$ isobar was then collected on the material, Fig A2. As expected, the material exhibits a two-step adsorption profile due to the cooperative mechanism and the steric bulk of the 2-ampd which prevents the $CO_2$ insertion from occurring in a concerted step.[9] Satisfyingly, the material captures 3.5 mmol/g of $CO_2$ at 35 °C (3.63 mmol/g theoretical maximum assuming 1 $CO_2$:1 diamine).

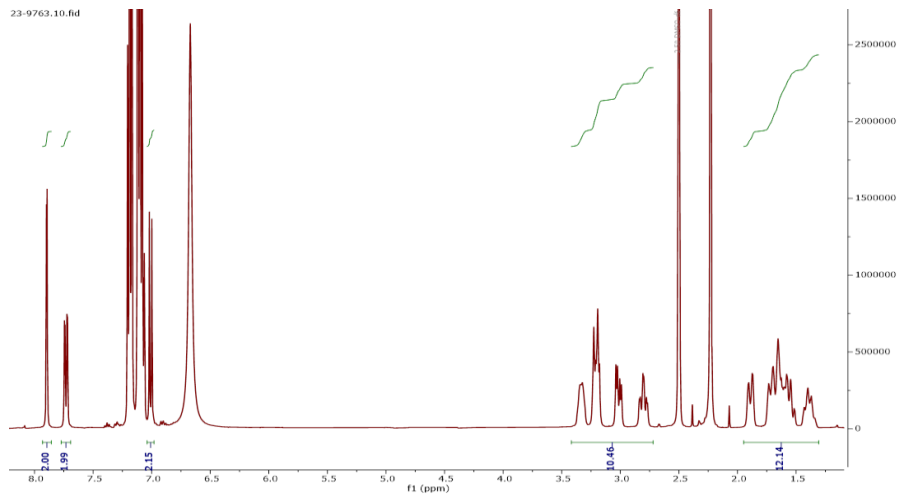

**Figure A1:** 1H NMR spectrum of a digested sample of MOF

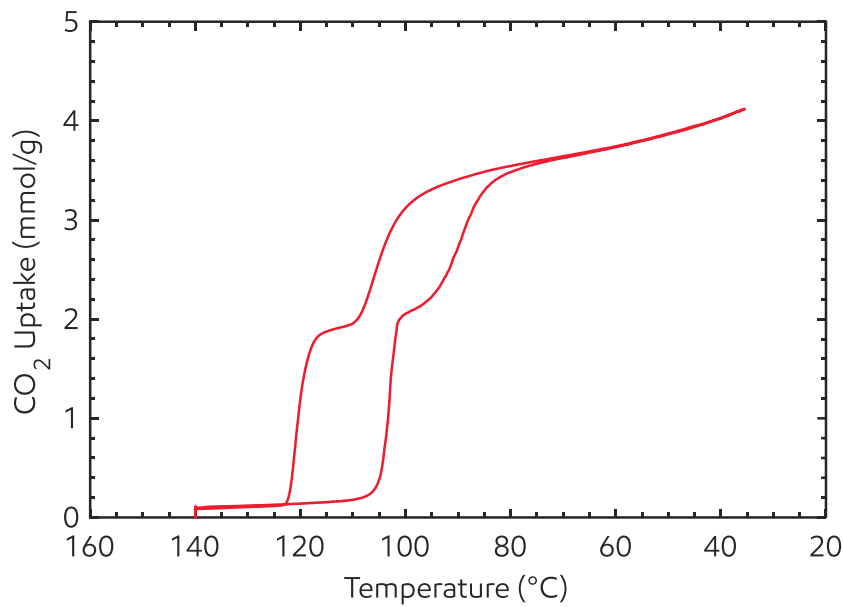

**Figure A2:** CO2 isobar of the material at a partial pressure of 100 kPa. Right curve gives adsorption branch, and left curve gives desorption branch

To gauge the material's propensity to adsorb $CO_2$ under DAC conditions, the material was exposed sequentially to dry 400 ppm $CO_2$ at 75 °C, 50 °C, and 25 °C followed by desorption at 140 °C between each adsorption step (Fig. S3). Due to the comparatively low enthalpy of adsorption of $CO_2$ on 2-ampd versus other amines, negligible adsorption of CO2 occurred at 75 and 50 °C. At 25 °C, however, the material steadily exhibits an increase in mass, attributed to the slow kinetics of $CO_2$ adsorption for this material under dry conditions.[9] The material was able to adsorb 0.54 mmol/g of $CO_2$ after an 8 hour exposure at 25 °C, however, the material had clearly not reached its equilibrium capacity under these conditions.

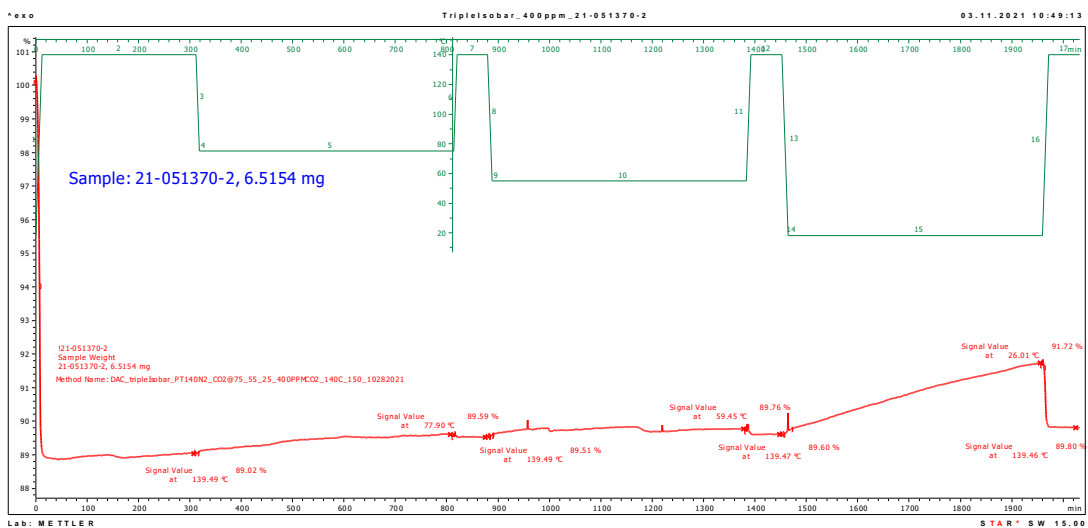

**Figure A3:** Dry $CO_2$ adsorption capacity at 75, 50, and 25 °C. A heat treatment step (140 °C under flowing $N_2$) was used to desorb $CO_2$ in between each adsorption step.

Acknowledging that many amines behave differently in the presence of water[18], we sought out to better characterize the material's $CO_2$ adsorption performance in the presence of water and under more realistic DAC conditions. To do this, we developed a humid TGA test (Fig. A4) in which the material was first exposed to a stream of $H_2O$ and $N_2$ at 22 °C with a partial pressure of $H_2O$ of roughly 1.59 kPa (60% relative humidity). The mass of the material increased quickly due to the adsorption of water and stabilized at 15.3 mmol/g. While maintaining the same relative humidity, the stream was changed to include 400 ppm $CO_2$ at which point the material increased in mass again attributed mostly to the adsorption of $CO_2$. However, we cannot preclude the presence of the ammonium carbamate sites changing the water adsorption behavior of the material and therefore cannot accurately assess the equilibrium $CO_2$ capacity of the material in mixed water/$CO_2$ streams using gravimetric techniques. To better estimate the amount of $CO_2$ adsorption, the stream was switched to dry 400 ppm $CO_2$ and the resulting mass loss was assumed to be purely water while the retained mass was assumed to be attributed to $CO_2$. Surprisingly, these data indicated the material had a $CO_2$ capacity of 2.57 mmol/g, a substantial increase from the expected equilibrium capacity of extrapolated from dry $CO_2$ isotherms. To be more confident in the equilibrium capacities from the gravimetric data, we sought out the measurement of multi-component breakthrough capacities where both water and $CO_2$ adsorption phenomenon can be more easily differentiated.

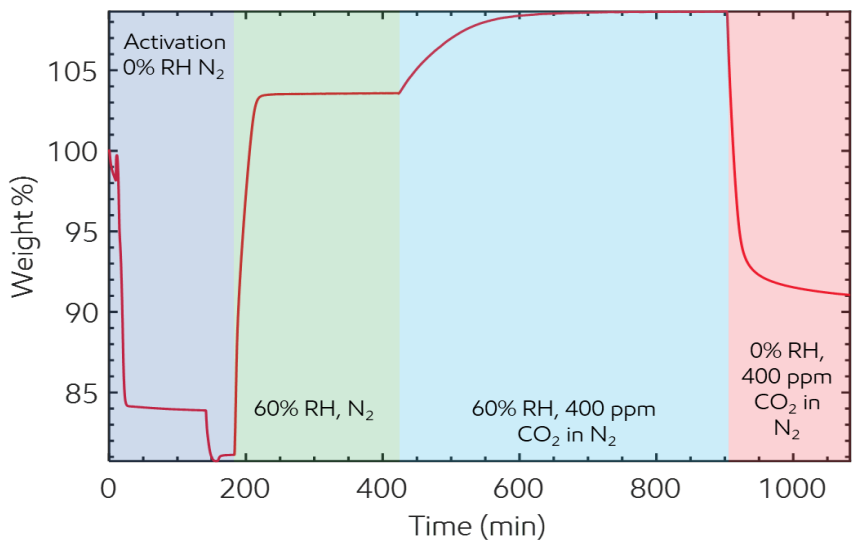

**Figure A4:** Adsorption behavior of MOF using a TGA. The material was first activated at 140 °C under a flowing stream of $N_2$. The sample was cooled down to 22 °C and exposed to 60% RH stream in $N_2$ (green), 60% RH and 400 ppm $CO_2$ ba. $N_2$ (light blue), followed by a dry 400 ppm $CO_2$ stream (red)

**References:**

1. Zhu, X.; Xie, W.; Wu, J.; Miao, Y.; Xiang, C.; Chen, C.; Ge, B.; Gan, Z.; Yang, F.; Zhang, M.; O'Hare, D.; Li, J.; Ge, T.; Wang, R., Recent advances in direct air capture by adsorption. *Chemical Society Reviews* **2022,** *51* (15), 6574-6651.
2. Sabatino, F.; Grimm, A.; Gallucci, F.; van Sint Annaland, M.; Kramer, G. J.; Gazzani, M., A comparative energy and costs assessment and optimization for direct air capture technologies. *Joule* **2021,** *5* (8), 2047-2076.
3. Keith, D. W.; Holmes, G.; St. Angelo, D.; Heidel, K., A Process for Capturing $CO_2$ from the Atmosphere. *Joule* **2018,** *2* (8), 1573-1594.
4. Bose, S.; Sengupta, D.; Rayder, T. M.; Wang, X.; Kirlikovali, K. O.; Sekizkardes, A. K.; Islamoglu, T.; Farha, O. K., Challenges and Opportunities: Metal–Organic Frameworks for Direct Air Capture. *Advanced Functional Materials* **2023**, 2307478.
5. Sharifzadeh, Z.; Morsali, A., Amine-Functionalized Metal-Organic Frameworks: from Synthetic Design to Scrutiny in Application. *Coordination Chemistry Reviews* **2022,** *459*, 214445.
6. Planas, N.; Dzubak, A. L.; Poloni, R.; Lin, L.-C.; McManus, A.; McDonald, T. M.; Neaton, J. B.; Long, J. R.; Smit, B.; Gagliardi, L., The mechanism of carbon dioxide adsorption in an alkylamine-functionalized metal–organic framework. *Journal of the American Chemical Society* **2013,** *135* (20), 7402-7405.
7. McDonald, T. M.; Lee, W. R.; Mason, J. A.; Wiers, B. M.; Hong, C. S.; Long, J. R., Capture of carbon dioxide from air and flue gas in the alkylamine-appended metal–organic framework mmen-Mg2 (dobpdc). *Journal of the American Chemical Society* **2012,** *134* (16), 7056-7065.
8. McDonald, T. M.; Mason, J. A.; Kong, X.; Bloch, E. D.; Gygi, D.; Dani, A.; Crocella, V.; Giordanino, F.; Odoh, S. O.; Drisdell, W. S., Cooperative insertion of $CO_2$ in diamine-appended metal-organic frameworks. *Nature* **2015,** *519* (7543), 303-308.
9. Siegelman, R. L.; Milner, P. J.; Forse, A. C.; Lee, J.-H.; Colwell, K. A.; Neaton, J. B.; Reimer, J. A.; Weston, S. C.; Long, J. R., Water Enables Efficient $CO_2$ Capture from Natural Gas Flue Emissions in an Oxidation-Resistant Diamine-Appended Metal–Organic Framework. *Journal of the American Chemical Society* **2019,** *141* (33), 13171-13186.
10. Holmes, H. E.; Ghosh, S.; Li, C.; Kalyanaraman, J.; Realff, M. J.; Weston, S. C.; Lively, R. P., Optimum relative humidity enhances $CO_2$ uptake in diamine-appended M2 (dobpdc). *Chemical Engineering Journal* **2023,** *477*, 147119.
11. Darunte, L. A.; Sen, T.; Bhawanani, C.; Walton, K. S.; Sholl, D. S.; Realff, M. J.; Jones, C. W., Moving beyond adsorption capacity in design of adsorbents for $CO_2$ capture from ultradilute feeds: kinetics of $CO_2$ adsorption in materials with stepped isotherms. *Industrial & Engineering Chemistry Research* **2018,** *58* (1), 366-377.
12. Kundu, J.; Stilck, J. F.; Lee, J.-H.; Neaton, J. B.; Prendergast, D.; Whitelam, S., Cooperative gas adsorption without a phase transition in metal-organic frameworks. *Physical review letters* **2018,** *121* (1), 015701.
13. Marshall, B. D., A Cluster Based Cooperative Kinetic Model for $CO_2$ Adsorption on Amine Functionalized Metal–Organic Frameworks. *Industrial & Engineering Chemistry Research* **2022,** *61* (49), 18138-18145.
14. Martell, J. D.; Milner, P. J.; Siegelman, R. L.; Long, J. R., Kinetics of cooperative CO 2 adsorption in diamine-appended variants of the metal–organic framework Mg 2 (dobpdc). *Chemical science* **2020,** *11* (25), 6457-6471.
15. Weston, S. C.; Abney, C. W.; Falkowski, J. M.; Ivashko, A., US20220176343A1. *US Patent Application* **2022**.

16.     Sircar, S.; Hufton, J., Why does the linear driving force model for adsorption kinetics work? *Adsorption* **2000,** *6* (2), 137-147.
17.     Lin, J.-B.; Nguyen, T. T. T.; Vaidhyanathan, R.; Burner, J.; Taylor, J. M.; Durekova, H.; Akhtar, F.; Mah, R. K.; Ghaffari-Nik, O.; Marx, S.; Fylstra, N.; Iremonger, S. S.; Dawson, K. W.; Sarkar, P.; Hovington, P.; Rajendran, A.; Woo, T. K.; Shimizu, G. K. H., A scalable metal-organic framework as a durable physisorbent for carbon dioxide capture. *Science* **2021,** *374* (6574), 1464-1469.
18.     Kolle, J. M.; Fayaz, M.; Sayari, A., Understanding the effect of water on CO2 adsorption. *Chemical Reviews* **2021,** *121* (13), 7280-7345.